# Smart metastructure method for increasing $T_C$ of Bi(Pb)SrCaCuO high-temperature superconductors

**Honggang Chen, Yongbo Li, Mingzhong Wang, Guangyu Han, Miao Shi, Xiaopeng Zhao***

Smart Materials Laboratory, Department of Applied Physics, Northwestern Polytechnical University, Xi'an 710129, China;

2017100698@mail.nwpu.edu.cn (H.C.); lybo2010@foxmail.com (Y.L.); wangmingzhongsuper@163.com (M.W.); 2226131321@qq.com (G.H.); 2711792400@qq.com (M.S.)

*Corresponding author: Prof. Xiaopeng Zhao, E-mail: xpzhao@nwpu.edu.cn

**Abstract:** Improving the critical transition temperature ($T_C$) of Bi(Pb)SrCaCuO (B(P)SCCO) high-temperature superconductors is important, however, considerable challenges exist. In this study, on the basis of the metamaterial structure and the idea that the injecting energy will promote the formation of Cooper pairs, a smart meta-superconductor B(P)SCCO consisting of B(P)SCCO microparticles and $Y_2O_3$:$Eu^{3+}$+Ag or $Y_2O_3$:$Eu^{3+}$ luminophor was designed. In the applied electric field, the $Y_2O_3$:$Eu^{3+}$+Ag or $Y_2O_3$:$Eu^{3+}$ luminophor generates an electroluminescence (EL), thereby promoting the $T_C$ via EL energy injection. A series of $Y_2O_3$:$Eu^{3+}$+Ag topological luminophor-doped B(P)SCCO samples was prepared. Results showed that $Y_2O_3$:$Eu^{3+}$+Ag was dispersed around B(P)SCCO particles, forming a metastructure. Accordingly, the onset transition temperature ($T_{C,on}$) and zero resistance transition temperature ($T_{C,0}$) of B(P)SCCO increased. Meanwhile, the B(P)SCCO sample doped with 0.2 wt% $Y_2O_3$ or $Y_2O_3$:$Sm^{3+}$ nonluminous inhomogeneous phase was also prepared to further prove the influence of EL on the $T_C$ rather than the rare earth effect. Results indicated that the $T_C$ of the $Y_2O_3$ or $Y_2O_3$:$Sm^{3+}$ doping sample decreased. However, the $T_C$ of the 0.2 wt% $Y_2O_3$:$Eu^{3+}$+Ag or $Y_2O_3$:$Eu^{3+}$ luminophor-doped sample improved. This outcome further demonstrated that the smart metastructure method can improve the $T_C$ of B(P)SCCO.

## 1. Introduction

Improving the critical transition temperature ($T_C$) of superconductors is important; however, considerable challenges exist. In 2011, Cavalleri et al. used a mid-infrared femtosecond laser pulse to induce the transformation of $La_{1.675}Eu_{0.2}Sr_{0.125}CuO_4$ from a nonsuperconducting into a transient three-dimensional superconductor [1]. The behavior of transient superconducting transition in $La_{1.84}Sr_{0.16}CuO_4$, $YBa_2Cu_3O_{6.5}$, and $K_3C_{60}$ was observed by using similar experimental methods [2-5]. The said researchers reported that laser pulse causes lattice distortion and induces transient superconductivity. Since then, the use of light to change the superconducting properties of materials has been gradually recognized. Scientists discovered the high-temperature superconductor BiSrCaCuO (BSCCO) with a $T_C$ beyond 100 K in 1988 [6, 7]. BiSrCaCuO superconductors are promising materials for theory research and industrial applications due to their several advantages, such as low oxygen sensitivity, containing no rare earth, and high $T_C$ [6-9]. The BiSrCaCuO system consists of three superconducting phases with similar crystal structures, and its general formula can be written as $Bi_2Sr_2Ca_{n-1}Cu_nO_{2n+4}$, where $n$=1, 2 and 3, with corresponding superconducting phases of Bi-2201 ($T_C$=20 K), Bi-2212 ($T_C$=85 K) and Bi-2223 ($T_C$=110 K), respectively [10-17]. Pure Bi-2223 and Bi-2212 single phase are difficult to obtain because they are symbiotic with each other, especially when forming the Bi-2223 phase [18-21]. However, partial replacement of Bi by Pb can increase the volume content of the Bi-2223 phase, thereby making it easy to synthesize and increasing its stability [22, 23].

Although Bi-based superconductors are called high-temperature superconductors, its critical parameters (especially the superconducting transition temperature $T_C$) are still far from the large-scale practical application. So Bi-based superconductors should be modified to increase its superconducting transition temperature $T_C$. At present, a commonly used method is chemical doping, for example, doping with elements, such as Cs [24], Al [25], Ce [26], and Pb [22, 23] in a Bi-Sr-Ca-Cu-O system. However, this method exhibits no significant increase in the superconducting transition temperature $T_C$. Subsequently, nanomaterials have been introduced for doping, for example, doping with $Al_2O_3$ [27], $SnO_2$ [28], $ZrO_2$ [29], MgO [30], $MgCO_3$ [31], and $Ca_2B_2O_5$ [32]. However, the results are unsatisfactory because most dopants are unstable at high temperature and react with the superconductor. Therefore, a suitable material for doping should be determined to ensure the stability at a high temperature and the increase of $T_C$.

Metamaterial, a type of artificially structured composite material, is composed of the matrix material and its unit material. The metamaterial properties are not primarily dependent on the matrix material but on the artificial structure. Many special functions can be obtained through various artificial structures [33-35]. With the development of metamaterial, the use of the metamaterial concept to design superconductors and affecting its $T_C$ has been gradually recognized by scholars. In 2007, our group introduced inorganic electroluminescence (EL) material in superconductor to enhance the superconducting transition temperature through EL. Jiang et al. [36] first introduced uniformly distributed ZnO nano defects with a doping concentration of 1 wt% in Bi(Pb)SrCaCuO (B(P)SCCO) superconductors. The effects of different doping methods on the superconducting transition temperature and morphology of B(P)SCCO superconductors were investigated. The results of the standard four-probe method indicated that samples doped with ZnO EL material showed an evident performance belonging to high-temperature superconductor. However, the doping of ZnO EL materials caused a slight decrease of the B(P)SCCO superconducting transition temperature. Fundamentally, $Y_2O_3$:$Eu^{3+}$ phosphor is an excellent rare earth luminescent material because of its several advantages, such as high luminescence intensity, good monochromaticity, high quantum efficiency. And the preparation process of such material is simple, and the morphology is relatively easy to control. Moreover, the preparation of $Y_2O_3$:$Eu^{3+}$ into a $Y_2O_3$:$Eu^{3+}$+Ag topological luminophor can further improve the EL properties of $Y_2O_3$:$Eu^{3+}$ and have better stability in the environment [37-39]. Recently, Smolyaninov et al. [40-42] proposed that a superconducting metamaterial with an effective dielectric constant $\varepsilon_{eff} \approx 0$ may exhibit high transition temperature, and they confirmed their theory in experiment.

Our group recently selected traditional $MgB_2$ superconductor and constructed a smart meta-superconductor $MgB_2$ model based on the metamaterial structure. Smart meta-superconductor $MgB_2$ consists of the $MgB_2$ matrix and inhomogeneous phases, such as the

EL material $Y_2O_3$:$Eu^{3+}$ rods and different sizes of $Y_2O_3$:$Eu^{3+}$ or $YVO_4$:$Eu^{3+}$ sheets. The research results showed that the doping of EL materials increases the superconducting transition temperature of $MgB_2$. This increment is attributed to the EL materials that dispersed around $MgB_2$ particles. In the local electric field, the EL materials generate an EL. Therefore, the $T_C$ of $MgB_2$ is improved by EL [43-47].

We select $MgB_2$ superconductor to construct a smart meta-superconductor $MgB_2$ based on the metamaterial structure and electron-phonon interaction in traditional $MgB_2$ superconductor. The electrons are transformed into Cooper pairs via energy injection by doping the $Y_2O_3$:$Eu^{3+}$ EL material in $MgB_2$ superconductor, thereby enhancing $T_C$ of $MgB_2$ [45-47]. Many scientists believe that Cooper pairs are formed on the basis of the electromagnetic interaction of electron spin in the high-temperature superconductor B(P)SCCO[48, 49]. A smart meta-superconductor B(P)SCCO is proposed in this work, which consists of B(P)SCCO particles and $Y_2O_3$:$Eu^{3+}$+Ag or $Y_2O_3$:$Eu^{3+}$ luminophor to form a composite particle structure. The B(P)SCCO superconducting particles are used as the matrix material, and the $Y_2O_3$:$Eu^{3+}$+Ag or $Y_2O_3$:$Eu^{3+}$ luminophor inhomogeneous phase distributed around the B(P)SCCO particles. In the local electric field, the B(P)SCCO superconducting particles act as microelectrodes, which excite the EL of $Y_2O_3$:$Eu^{3+}$+Ag or $Y_2O_3$:$Eu^{3+}$ luminophor, and EL energy injection will promote the formation of Cooper pairs, thus changing the $T_C$ of B(P)SCCO. The $Y_2O_3$:$Eu^{3+}$+Ag topological luminophor-doped B(P)SCCO samples are prepared with different doping concentrations [50]. The results show that all samples have evident superconducting transition, and $Y_2O_3$:$Eu^{3+}$+Ag topological luminophor-doped improves the superconducting transition temperature of B(P)SCCO superconductor. The nonluminous inhomogeneous phases $Y_2O_3$ and $Y_2O_3$:$Sm^{3+}$ doping B(P)SCCO are prepared to further prove the effect of EL. The result shows that the nonluminous inhomogeneous phases $Y_2O_3$ and $Y_2O_3$:$Sm^{3+}$ doping decreases the superconducting transition temperature.

## 2. Model

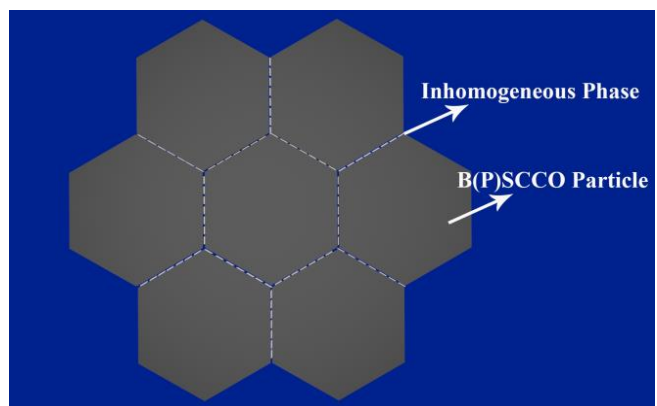

**Figure 1.** The model of the smart meta-superconductor B(P)SCCO

Figure 1 shows the microstructure model of the smart meta-superconductor B(P)SCCO based on the metamaterial structure. The black hexagons in this figure represent the B(P)SCCO superconducting particles, and the $Y_2O_3$:$Eu^{3+}$+Ag or $Y_2O_3$:$Eu^{3+}$ luminophor inhomogeneous phase is dispered around the B(P)SCCO particles, just like the discontinuous white ones in this figure. This model consists of B(P)SCCO superconducting particles and $Y_2O_3$:$Eu^{3+}$+Ag or $Y_2O_3$:$Eu^{3+}$ luminophor to form a composite particle structure. The B(P)SCCO superconducting particles are used as the matrix material, and the $Y_2O_3:Eu^{3+}$+Ag or $Y_2O_3:Eu^{3+}$ luminophor distributed around the B(P)SCCO particles are used as inhomogeneous phase dopants. When using a four-probe method in a liquid helium cryogenic system to measure the curve of the temperature dependence of resistivity (R–T) of the samples, the B(P)SCCO particles act as microelectrodes, which excite the EL of inhomogeneous phase EL materials, and EL energy injection will promote the formation of Cooper pairs. Thus the $T_C$ of B(P)SCCO will be improved by EL energy injection. Adjusting the applied electric field to control the EL of $Y_2O_3:Eu^{3+}$+Ag or $Y_2O_3:Eu^{3+}$ luminophor may alter the $T_C$ of this smart meta-superconductor, thereby achieving a smart meta-superconductor.

**3. Experiment**

3.1. Preparation of inhomogeneous phase dopants

The preparation process of the topological luminophor $Y_2O_3:Eu^{3+}$+Ag (marked as N1) was described in detail in literature 37. $Y_2O_3$ (marked as N2), $Y_2O_3:Sm^{3+}$ (marked as N3) nonluminous inhomogeneous phases and $Y_2O_3:Eu^{3+}$ (marked as N4) luminous inhomogeneous phase were obtained by changing the raw material.

3.2. Preparation of pure B(P)SCCO superconductor

A certain amount of raw material (with a purity of 99% or 99.99%) was weighed according to the molar ratio of $Bi_2O_3:PbO:SrCO_3:CaCO_3:CuO$=0.92:0.34:2.00:2.00:3.00. The powders were mixed and grounded, followed by ball milling for 20 h at a speed of 500 r/min in an appropriate amount of anhydrous ethanol. The slurry was then dried at 60 ℃ to obtain gray powder. The dried gray powder was placed in a tube furnace, kept at 830 ℃ for 10 h, cooled to room temperature and then grounded in an agate mortar. The process was repeated once to obtain B(P)SCCO calcined powder. The calcined powder was sufficiently grounded and then kept at 10 MPa for 10 min to form a pellet of 12 mm diameter and 2 mm thickness. Finally, the pellet was placed in a tube furnace at 830 ℃ for 10 h to obtain a pure B(P)SCCO sample [50].

3.3. Preparation of inhomogeneous phase doping B(P)SCCO superconductors

Inhomogeneous phase dopants and B(P)SCCO calcined powder were mixed in 20 mL of ethanol and stirred 20 min with a magnetic stirrer to form a suspension. The suspension was transferred into a culture dish after 20 min of sonication and dried in a vacuum drying oven at 60 ℃ for 4 h to obtain black powder. The black powder was then fully grounded and kept at 10 MPa for 10 min to form a pellet of 12 mm diameter and 2 mm thickness. Afterward, the pellet was placed in a tube furnace at 830 ℃ for 10 h to obtain the corresponding inhomogeneous phase doping B(P)SCCO sample [50]. We used two different purity raw materials to prepare nine types of doped samples, the contents and types of dopants in all samples are shown in Table 1.

**Table 1** Purity, dopant and doping concentration of all the samples

| Sample | A1 | A2 | A3 | A4 | A5 | A6 | B1 | B2 | B3 | B4 | B5 |
|---|---|---|---|---|---|---|---|---|---|---|---|
| Purity (%) | 99 | 99 | 99 | 99 | 99 | 99 | 99.99 | 99.99 | 99.99 | 99.99 | 99.99 |
| Dopant | None | N1 | N1 | N1 | N2 | N3 | None | N1 | N4 | N2 | N3 |
| Concentration (wt%) | 0 | 0.1 | 0.2 | 0.5 | 0.2 | 0.2 | 0 | 0.2 | 0.2 | 0.2 | 0.2 |

## 4. Results and discussion

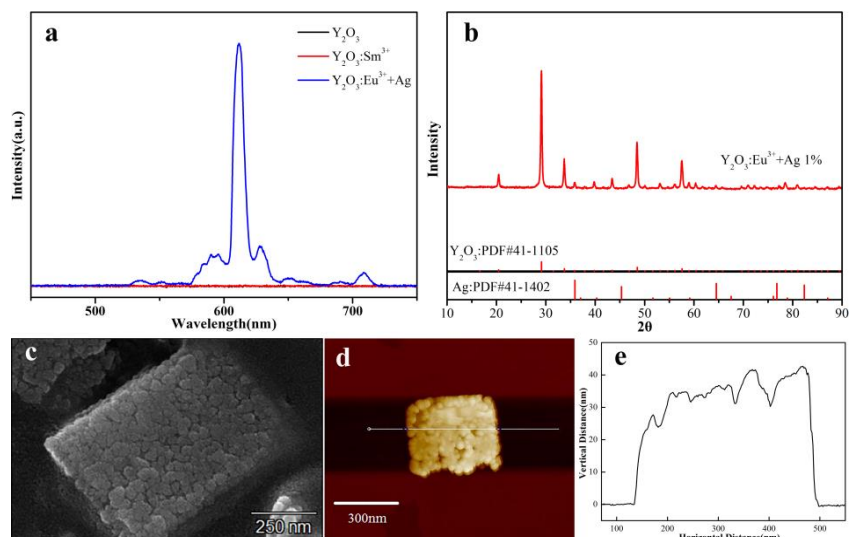

**Figure 2.** EL spectrum of different dopants (a); XRD (b), SEM (c), and AFM (d) image of the $Y_2O_3:Eu^{3+}+Ag$ topological luminophor (N1); (e) Height profile corresponding to the lines draw in d

In order to prepare a metastructure superconductor consisting of B(P)SCCO superconductor and $Y_2O_3:Eu^{3+}+Ag$ topological luminophor, we initially synthesized the $Y_2O_3:Eu^{3+}+Ag$ topological luminophor. Figure 2a shows the EL spectrum of the $Y_2O_3:Eu^{3+}+Ag$ topological luminophor. We also synthesized $Y_2O_3:Sm^{3+}$ and $Y_2O_3$ dopants to further demonstrate the effect of EL on metamaterial superconductor B(P)SCCO. Figure 2a also shows the EL spectrum of $Y_2O_3:Sm^{3+}$ and $Y_2O_3$. The spectrum shows that a strong peak centered at 613 nm, which corresponds to $Eu^{3+}$ ions typical of the transition from $^5D_0$ to $^7F_2$. The $Y_2O_3:Eu^{3+}$ system formed by the nonluminous $Y_2O_3$ and luminous center $Eu^{3+}$ ions is a strong luminophor. The EL intensity of $Y_2O_3:Eu^{3+}$ can be further enhanced by Ag doping, and the EL intensity of the $Y_2O_3:Eu^{3+}+Ag$ topological luminophor is considerably stronger than those of $Y_2O_3:Sm^{3+}$ and $Y_2O_3$. Figure 2b shows the X-ray diffraction (XRD) pattern of the $Y_2O_3:Eu^{3+}+Ag$ topological luminophor. The image indicates that the prepared $Y_2O_3:Eu^{3+}+Ag$ topological luminophor is pure $Y_2O_3$, and no other impurity phases are detected. The Eu and Ag are added in small amounts; thus, no evident diffraction peak is found in the XRD pattern. Figure 2c–e show the scanning electron microscopy (SEM) and atomic force microscopy (AFM) images of the $Y_2O_3:Eu^{3+}+Ag$ topological luminophor. The prepared $Y_2O_3:Eu^{3+}+Ag$ topological luminophor dopant is a flake structure with a size of 300×400 nm and a thickness of approximately 35 nm.

Figure 3 shows the XRD patterns of the pure B(P)SCCO sample (A1) and different inhomogeneous phase doping samples (A2, A3, A4, A5 and A6) prepared by solid-state sintering. The characteristic peaks of high-temperature phase Bi-2223 and low-temperature phase Bi-2212 are labeled by a rhombus and triangle, respectively. The peak positions and intensities of diffraction indicate that all samples comprise a mixture of high-temperature phase Bi-2223 and low-temperature phase Bi-2212, and no other impurity phases are detected. Besides, the addition of dopants has not introduced other impurity phases.

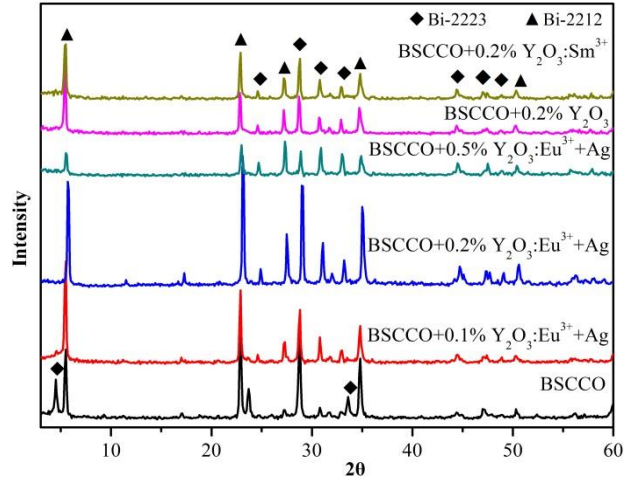

**Figure 3.** X-ray diffraction patterns of A1, A2, A3, A4, A5 and A6

In this study, all peaks of the Bi-2223 and Bi-2212 phase have been used for the calculation of the volume content of the phases. The volume contents of the high-temperature phase Bi-2223 and low-temperature phase Bi-2212 calculated using the following equations [51, 52] are listed in Table 2:

$$\mathrm{Bi2223(\%)} \approx \frac{\sum I(Bi2223)}{\sum I(Bi2223) + \sum I(Bi2212)} \times 100\%,$$

$$\mathrm{Bi2212(\%)} \approx \frac{\sum I(Bi2212)}{\sum I(Bi2223) + \sum I(Bi2212)} \times 100\%,$$

where $I$ is the intensity of the Bi-2223 and Bi-2212 phase in the XRD pattern (Figure 3). Table 2 illustrates that the volume contents of the high-temperature phase Bi-2223 in the doped samples are slightly decreased, and that of the low-temperature phase Bi-2212 in all prepared samples are more than 50%, which result in a wide superconducting transition width and a low $T_C$.

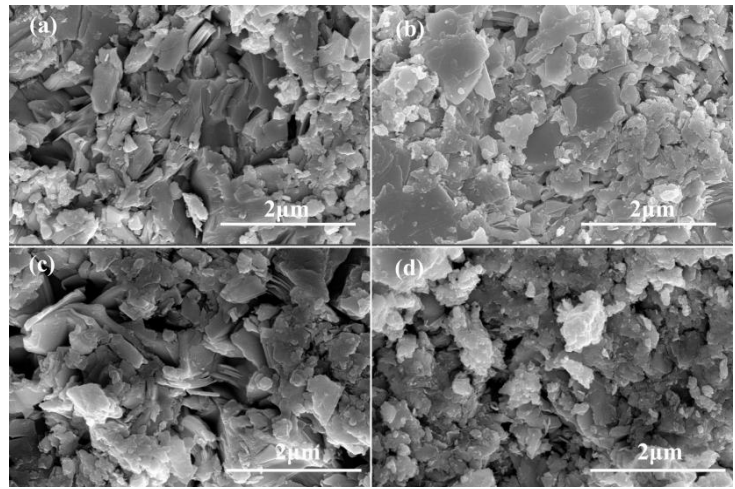

**Figure 4.** SEM images of A1 (a), A3 (b), A5 (c), and A6 (d)

Figure 4a–d show the SEM images of the pure B(P)SCCO sample (A1), the 0.2 wt% $Y_2O_3$:$Eu^{3+}$+Ag doped B(P)SCCO sample (A3), the 0.2 wt% $Y_2O_3$ doped sample (A5), and the 0.2 wt% $Y_2O_3$:$Sm^{3+}$ doped sample (A6), respectively. The images manifest that all the prepared samples are irregular block structure with a particle size of less than 2 μm. The addition of the dopants does not affect the microstructure of B(P)SCCO. Since the content of the $Y_2O_3$:$Eu^{3+}$+Ag topological luminophor in the sample is small, no flake inhomogeneous phase dopants are found in the doped samples, and no $Y_2O_3$ peaks are detected in the XRD pattern of the doped samples, so an elemental analysis and X-ray photoelectron spectrometric (XPS) were performed. Figure 5 illustrates the distribution of certain chemical elements. The top left corner of each figure shows the corresponding element. Figure 6a, 6b show the XPS spectra of $Y_2O_3$:$Eu^{3+}$+Ag topological luminophor and the 0.5 wt% $Y_2O_3$:$Eu^{3+}$+Ag doped B(P)SCCO sample (A4), respectively. The peaks of Eu 3d and Ag 3d were observed in the XPS spectra of $Y_2O_3$:$Eu^{3+}$+Ag topological luminophor (Fig. 6a), indicating that Eu and Ag were present in the topological luminophor [37]. And Y was observed in SEM/EDS (Fig. 5) and the XPS (Fig. 6b) spectra of topological luminophor doped sample, indicating the presence of the compound $Y_2O_3$ in the doped sample. Therefore, $Y_2O_3$:$Eu^{3+}$+Ag topological luminophor existed in the doped sample, and $Y_2O_3$:$Eu^{3+}$+Ag topological luminophor distributed around the B(P)SCCO particles.

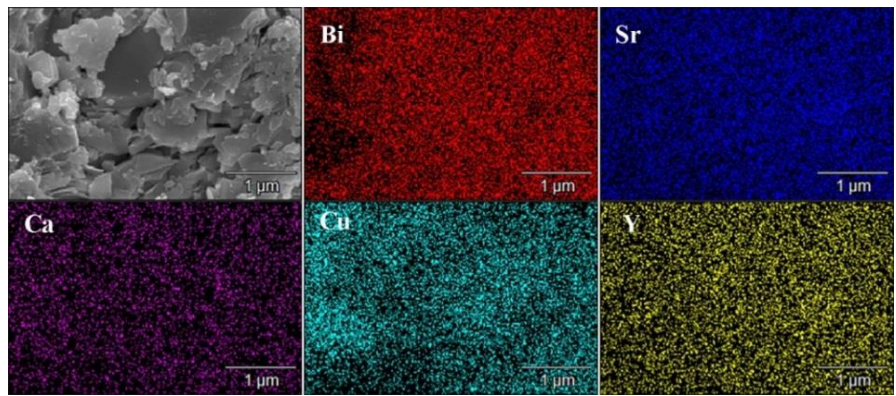

**Figure 5.** SEM image and Chemical element distribution map of A4

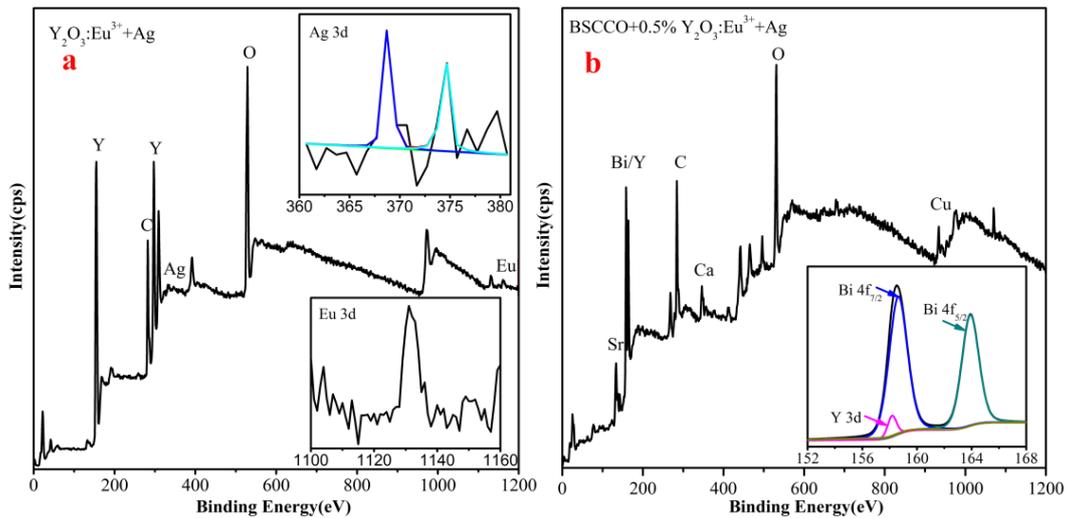

**Figure 6a.** XPS spectra of $Y_2O_3$:$Eu^{3+}$+Ag, the insets show the magnified Ag 3d and Eu 3d spectra. **6b.** XPS spectra of A4, the inset show the magnified Bi 4f and Y 3d spectra

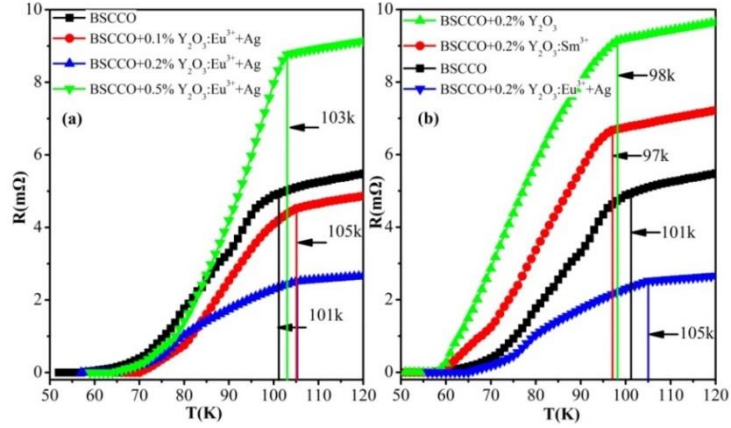

**Figure 7.** Temperature-dependent resistivity curves of A1, A2, A3, A4, A5 and A6

Figure 7a presents the *R–T* curve of the pure B(P)SCCO (A1), and B(P)SCCO doped with 0.1 wt% $Y_2O_3$:$Eu^{3+}$+Ag (A2), 0.2 wt% $Y_2O_3$:$Eu^{3+}$+Ag (A3), and 0.5 wt% $Y_2O_3$:$Eu^{3+}$+Ag (A4) with a test current of 100 mA. The electrical resistivity is measured using the standard four-probe method. All prepared samples show a superconducting transition between 50 K and 120 K. The two characteristic temperatures, namely, onset transition temperature $T_{C,on}$ and zero–resistivity transition temperature $T_{C,0}$, on each *R–T* curve are discussed. $T_{C,on}$ and $T_{C,0}$ are defined by generally accepted standards in literatures 22, 53. The resistance temperature (*R–T*) curve exhibits metallic-like behavior between $T_{C,on}$ and room temperature. $T_{C,on}$ is the temperature at which the *R–T* curve deviates from linear behavior during cooling process, and the slope of the *R–T* curve changes significantly before and after this point. $T_{C,0}$ is the temperature at which the resistance just completely drops to zero. The black curve shows the *R–T* curve of pure B(P)SCCO. The $T_{C,0}$ and $T_{C,on}$ of pure B(P)SCCO are 57 K and 101 K, respectively. The low transition temperature may be due to the low sintering temperature, extremely short sintering time, and insufficient grinding of each sintering. The transition temperatures $T_{C,0}$ and $T_{C,on}$ of the $Y_2O_3$:$Eu^{3+}$+Ag topological luminophor doping samples exhibit a increase compared with the pure B(P)SCCO sample, which maybe due to the $Y_2O_3$:$Eu^{3+}$+Ag topological luminophor distributed around B(P)SCCO particles to form a metamaterial structure with a special response, when testing the *R–T* curve of the sample, the B(P)SCCO particles act as microelectrodes, and the inhomogeneous phase EL materials would generate an EL, thereby the $T_C$ of B(P)SCCO can be improved by EL energy injection. The transition temperatures are listed in Table 2. And the amplitude of the increase in transition temperature decreases with the increase of the doping concentration.

In order to further confirm whether the increase in transition temperature is the effect of EL or rare earth, the samples doped with 0.2 wt% $Y_2O_3$ and $Y_2O_3$:$Sm^{3+}$ were prepared. Figure 7b depicts the *R–T* curve of the pure B(P)SCCO (A1), and B(P)SCCO doped with 0.2 wt% $Y_2O_3$:$Eu^{3+}$+Ag (A3), 0.2 wt% $Y_2O_3$ (A5) and 0.2 wt% $Y_2O_3$:$Sm^{3+}$ (A6) with a test current of 100 mA. It can be seen that the $T_{C,0}$ and $T_{C,on}$ of the 0.2 wt% $Y_2O_3$:$Eu^{3+}$+Ag topological luminophor doped sample show an obvious increase compared with the pure B(P)SCCO sample, and those of B(P)SCCO doped with 0.2 wt% $Y_2O_3$:$Eu^{3+}$+Ag topological luminophor are increased by 8 K and 4 K, respectively. However the $T_C$ of B(P)SCCO doped with 0.2 wt% $Y_2O_3$ and 0.2 wt% $Y_2O_3$:$Sm^{3+}$ nonluminous inhomogeneous phases show a decrease. This finding confirms that the increase of transition temperature is the effect of EL rather than the influence of rare earth elements.

**Table 2** Summary of the volume content and critical temperature of A1, A2, A3, A4, A5 and A6

| Sample | Bi-2223 (%) | Bi-2212 (%) | 100mA/K |
|--------|-------------|-------------|---------|
| A1 | 46.8 | 53.2 | 57-101 |
| A2 | 46.2 | 53.8 | 70-105 |
| A3 | 45.1 | 54.9 | 65-105 |
| A4 | 43.4 | 56.6 | 64-103 |
| A5 | 44.8 | 55.2 | 58-98 |
| A6 | 44.5 | 55.5 | 58-97 |

We found that the purity of raw materials and test status are the root causes of the poor quality of the samples and transition curves. So we changed the purity of raw materials from 99% to 99.99%, the quality of the sample improved, the volume fraction of Bi-2223 increased from 46.8% to 74.3%, and the volume fraction of Bi-2212 decreased from 53.2% to 25.7%. $T_{C,0}$ increased from 57 K to 67 K, and $T_{C,on}$ increased from 101 K to 109 K. Figure 8 depict the $R$–$T$ curve of the pure B(P)SCCO with a raw materials purity of 99.99% (B1), and B(P)SCCO doped with 0.2 wt% $Y_2O_3$:$Eu^{3+}$+Ag (B2), $Y_2O_3$:$Eu^{3+}$ (B3), $Y_2O_3$ (B4) and $Y_2O_3$:$Sm^{3+}$ (B5) at different test currents. The transition temperatures are listed in Table 3. With the test current $I$ decreases from 100 mA to 0.1 mA, $T_{C,0}$ of pure B(P)SCCO increases from 67 K to 89 K, and $T_{C,on}$ remains unchanged ($T_{C,on}$=109 K). When $I$=1 and 0.1 mA, the transition temperature of pure B(P)SCCO is 81-109 and 89-109 K, respectively, and the transition width is small. At the same time, we still found that doping of $Y_2O_3$ and $Y_2O_3$:$Sm^{3+}$ nonluminous dopants reduces the transition temperature of B(P)SCCO, however $Y_2O_3$:$Eu^{3+}$ and $Y_2O_3$:$Eu^{3+}$+Ag luminous inhomogeneous phases doping increases the transition temperature of B(P)SCCO by 2~3 K.

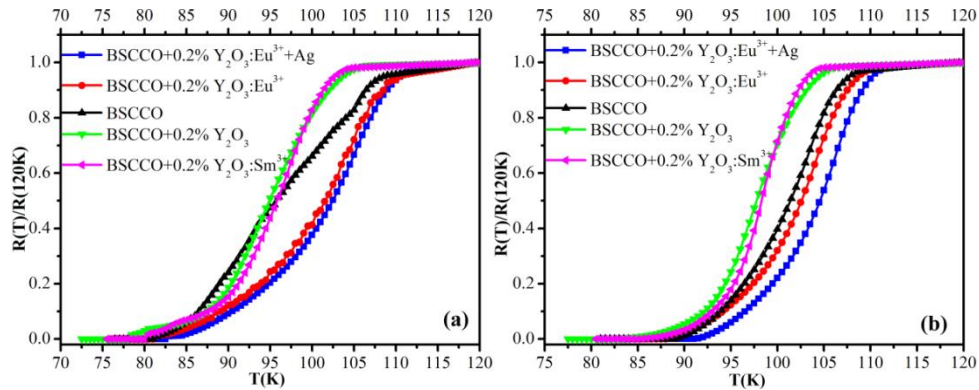

**Figure 8.** Temperature-dependent resistivity curves of B1, B2, B3, B4 and B5 at (a) 1 mA, (b) 0.1 mA

**Table 3** Summary of the volume content and critical temperature of B1, B2, B3, B4 and B5

| Sample | Bi-2223 (%) | Bi-2212 (%) | 100mA/K | 10mA/K | 1mA/K | 0.1mA/K |
|--------|-------------|-------------|---------|--------|-------|---------|
| B1 | 74.3 | 25.7 | 67-109 | 78-109 | 81-109 | 89-109 |
| B2 | 74.2 | 25.8 |  | 80-112 | 83-112 | 91-112 |
| B3 | 74.5 | 25.5 |  | 78-111 | 81-111 | 88-111 |
| B4 | 74.4 | 25.6 |  | 74-107 | 78-107 | 83-107 |
| B5 | 74.7 | 25.4 |  | 76-105 | 80-105 | 85-105 |

This experiment indicates that the $T_C$ of B(P)SCCO increased by doping with $Y_2O_3$:$Eu^{3+}$+Ag or $Y_2O_3$:$Eu^{3+}$ luminophor. However, Jiang et al. [36] found that the $T_C$ of B(P)SCCO doped with 1 wt% ZnO EL material decreased compared with pure B(P)SCCO in 2007. Two experiments present different results, which may be explained by the following reasons: The microstructure of ZnO dopants are spherical or ellipsoid, whereas the $Y_2O_3$:$Eu^{3+}$+Ag or $Y_2O_3$:$Eu^{3+}$ luminophor used in this experiment exhibits a flake structure, this structure can further improve the dispersion and connectivity, thereby making the dispersion more uniform, better connectivity, and more consistent with the proposed model. And ZnO EL intensity is extremely weak, the EL intensity of $Y_2O_3$:$Eu^{3+}$+Ag or $Y_2O_3$:$Eu^{3+}$ luminophor in this experiment is considerably stronger than that of ZnO. Meanwhile, the doping concentration of ZnO is extremely high.

The superconducting mechanism of the traditional $MgB_2$ superconductor is the interaction of electron and phonon, and the transformation of electrons into Cooper pairs can be enhanced via EL energy injection by doping with $Y_2O_3$:$Eu^{3+}$ EL materials in $MgB_2$ superconductor, thereby enhancing $T_C$ of $MgB_2$ [45-47]. In this experiment, the $T_{C,0}$ and $T_{C,on}$ of high-temperature superconductor B(P)SCCO increased by doping with $Y_2O_3$:$Eu^{3+}$+Ag or $Y_2O_3$:$Eu^{3+}$ luminophor, which maybe due to Cooper pairs are formed on the basis of the electromagnetic interaction of electron spin in the high-temperature B(P)SCCO superconductor, and the EL energy injection of $Y_2O_3$:$Eu^{3+}$+Ag or $Y_2O_3$:$Eu^{3+}$ luminophor promotes the formation of Cooper pairs, thus the $T_C$ of B(P)SCCO increased. Of course, this increase may also be related to the creation of charge carriers by doping the conduction bands [54-57]. The mechanism for enhancing the $T_C$ is unclear and requires further exploration.

## 5. Conclusion

Based on the idea that injecting energy will promote the formation of Cooper pairs, a smart meta-superconductor B(P)SCCO is constructed according to the method of metastructure, which consists of B(P)SCCO particles and $Y_2O_3$:$Eu^{3+}$+Ag or $Y_2O_3$:$Eu^{3+}$ luminophor to form a composite particle structure. In the local electric field, the B(P)SCCO superconducting particles act as microelectrodes, which stimulate the EL of $Y_2O_3$:$Eu^{3+}$+Ag or $Y_2O_3$:$Eu^{3+}$ luminophor, thereby improving the $T_C$ by EL energy injection. A series of B(P)SCCO samples doped with different amounts of $Y_2O_3$:$Eu^{3+}$+Ag topological luminophor was experimentally prepared. The volume content of the low-temperature phase Bi-2212 in all prepared samples is more than 50% by XRD characterization, which result in a wide superconducting transition width and a low $T_C$. The SEM and distribution of chemical elements indicated that the prepared samples are randomly oriented and exhibit an irregular blocky structure. The addition of the dopants does not affect the formation and microstructure of B(P)SCCO. The inhomogeneous phase dopants are distributed around the B(P)SCCO particles and formed a metastructure, so the effectiveness of EL in improving the $T_{C,0}$ and $T_{C,on}$ of B(P)SCCO can be fully reflected. At the same time, we prepared B(P)SCCO samples doped with 0.2 wt% $Y_2O_3$ and 0.2 wt% $Y_2O_3$:$Sm^{3+}$ nonluminous inhomogeneous phase to further demonstrate the influence of EL on $T_C$ rather than the rare earth effect. The results show that the $T_C$ of $Y_2O_3$ or $Y_2O_3$:$Sm^{3+}$ nonluminous inhomogeneous phase doping sample decreases. However, the $T_{C,0}$ and $T_{C,on}$ of the 0.2 wt% $Y_2O_3$:$Eu^{3+}$+Ag topological luminophor-doped sample increase. When we change the purity of the raw material to obtain the new high-purity samples and use the standard low current to test, the transition temperature and transition width of our newly prepared samples are in good agreement with the results in the literature. At this time, we still find that doping of $Y_2O_3$ and $Y_2O_3$:$Sm^{3+}$ nonluminous inhomogeneous phases reduces the $T_C$ of B(P)SCCO. while $Y_2O_3$:$Eu^{3+}$ and $Y_2O_3$:$Eu^{3+}$+Ag luminous inhomogeneous phases doping increases the $T_C$ of B(P)SCCO. This outcome may be that the $Y_2O_3$:$Eu^{3+}$+Ag or $Y_2O_3$:$Eu^{3+}$ luminophor generates an EL under the action of an applied electric field, thereby improving the $T_C$ of B(P)SCCO via energy injection. It's significant to improve the $T_C$ of high-temperature superconductor B(P)SCCO, in this study we construct a smart meta-superconductor B(P)SCCO to promote the formation of Cooper pairs via EL energy injection, this provides a new idea for improving the $T_C$ and practical application of high-temperature superconductors.

**Acknowledgements**

This work was supported by the National Natural Science Foundation of China for Distinguished Young Scholar under Grant No. 50025207.

**Author Contributions**

X.Z. conceived and led the project; H.C. and X.Z. designed the experiments; H.C., Y.L., M.W., G.H. and M.S. performed the experiments and characterized the samples; all authors discussed and analyzed the results; H.C. wrote the paper with input from all co-authors; X.Z. and H.C. discussed the results and revised the manuscript.

**Additional Information**
**Conflicts of Interest:** The authors declare no conflict of interest.